\documentstyle[aps,epsf,preprint,graphicx,prl]{revtex}

\begin{document} 
\draft

\def\beq{\begin{equation}}
\def\eeq{\end{equation}}
\def\beqn{\begin{eqnarray}}
\def\eeqn{\end{eqnarray}}
\def\btimes {\mbox{\boldmath $\times$}}
\def\bbox {\mbox{\boldmath $\box$}}
\def\bvarphi {\mbox{\boldmath $\varphi$}}
\def\ed{\end{document}}

\def\veps {{\varepsilon}}
\def\I {{\bf I}}
\def\II {{\bf II}}
\def\III {{\bf III}}
\def\IV {{\bf IV}}
\def\V {{\bf V}}
\def\VI {{\bf VI}}
\def\J {{\bf J}}
\def\H {{\bf H}}
\def\E {{\bf E}}
\def\1 {{\bf 1}}
\def\2 {{\bf 2}}
\def\3 {{\bf 3}}
\def\P {{\bf P}}
\def\r {{\bf r}}
\def\k {{\bf k}}
\def\p {{\bf p}}
\def\n {{\bf n}}
\def\A {{\bf A}}
\def\bv {{\bf v}} 
\def\AAN {$\!\!\!$ A$^{^{\!\!\!\!\! {\tiny {\circ}}}}$}
\def\aaN {$\!\!$ a$^{^{\!\!\!\! {\tiny {\circ}}}}$}

\title{Zoo of quantum phases and excitations of cold bosonic atoms\\ in optical lattices}

\author{Ofir E. Alon\footnote{E-mail: ofir@tc.pci.uni-heidelberg.de}, Alexej I. Streltsov and Lorenz S. Cederbaum}
\address{Theoretische Chemie, Physikalisch-Chemisches Institut, Universit\"at Heidelberg,\\
Im Neuenheimer Feld 229, D-69120 Heidelberg, Germany}

\maketitle

\begin{abstract}
Quantum phases and phase transitions of
weakly- to strongly-interacting bosonic atoms in deep
to shallow optical lattices are described by a 
{\it single multi-orbital mean-field approach in real space}. 
For weakly-interacting bosons in 1D,
the critical value of the superfluid to Mott insulator (MI) transition found 
is in excellent agreement with {\it many-body} treatments of the Bose-Hubbard model.
For strongly-interacting bosons,
 (i) additional MI phases appear, for which two (or more)
atoms residing in {\it each site} undergo a Tonks-Girardeau-like transition and 
localize and (ii) on-site excitation becomes the excitation lowest in energy. 
Experimental implications are discussed.
\end{abstract}

Loading and manipulating cold bosonic atoms in optical lattices
is a fascinating, rapidly growing branch of cold-atom physics, 
see, e.g., Refs.~\cite{IB1_nature,TE1_PRL,IB2_nature}.
In a pioneering experiment, Griener {\it et al.} \cite{IB1_nature}
have demonstrated the quantum phase transition
from the superfluid (SF) to Mott insulator (MI) phase 
of a cold gas of $^{87}$Rb atoms trapped by a three-dimensional simple-cubic-type optical lattice.
More recently, the SF to MI transition has been demonstrated in an effective 
one-dimensional (1D) optical lattice \cite{TE1_PRL}.
In the SF phase, which has no excitation gap, 
atoms are free to move throughout the lattice and
are associated with a coherent state of matter. 
The MI phase amounts for commensurate filling of the optical lattice 
and has an excitation gap associated with moving an atom from one site to 
an occupied neighboring site.
The SF to MI transition (in deep optical lattices) 
involves {\it weakly-interacting} bosons and is well described by the 
Bose-Hubbard model \cite{Fisher_PRB,Jaksch1_PRL,Zwerger1_PRL}
which assumes all bosons to occupy the lowest band of the lattice.
Recently, the so-called Tonks-Girardeau gas, i.e., the {\it strongly-interacting} regime, 
was realized in an optical lattice \cite{IB2_nature}.
Generally, as the interaction between atoms increases --  in deep as well as in shallow optical lattices -- 
additional possibilities open up 
for the trapped cold atoms which can now occupy higher bands.

Our purpose in this letter is to explore quantum phases and excitations
of cold bosonic atoms in optical lattices not accounted for so far.
We will present an approach which is able to treat {\it weakly- to  
strongly-interacting bosons in the entire range of deep to shallow optical lattices}.
To peruse the above, we would like to describe cold bosonic atoms in optical lattices directly in {\it real-space},
i.e., to provide their spatial wavefunction.
Specifically,
we will adopt our recently introduced {\it multi-orbital mean-field} \cite{LA_PLA,LA_ALN_PRA}
to cold bosonic atoms in optical lattices.
As we shall see below,
already for the standard case of weakly-interacting atoms we found a promising and intriguing result
concerning the critical value of the SF to MI transition in 1D. 
Although our approach is {\it mean-field}, it finds this value to be 
$3.855(7)$,
in excellent agreement with density matrix renormalization group and other {\it many-body} calculations,
see Ref.~\cite{Zwerger2_EPL} and references therein.

Our starting point is the many-body Hamiltonian describing $N$ bosons in an optical lattice,
$\hat H =  \sum_{i=1}^{N} \left[ \hat T(\r_i) +  V(\r_i) \right] +
 \sum_{i>j=1}^N U(\r_i-\r_j)$.
Here, $\r_i$ 
is the coordinate of the $i$-th particle,
$\hat T(\r_i)$ and $V(\r_i)$ stand for the kinetic energy and optical lattice potential, respectively,
and  $U(\r_i-\r_j)$ 
describes the pairwise contact interaction between the $i$-th and $j$-th atoms.

As mentioned above, we are going to obtain a real-space, wavefunction picture
of the quantum state of cold atoms in the optical lattice.
How are we going to achieve that?
To this end, we attach an {\it orbital} to each of the $N$ atoms.
The {\it simplest} choice is the Gross-Pitaevskii approach, 
for which {\it all} bosons reside in the {\it same} orbital.
There can be, however, many other situations for bosons \cite{LA_PLA}.
Generally, we may take $n_1$ bosons to reside in one orbital, $\varphi_1(\r)$,
$n_2$ bosons to reside in a second orbital, $\varphi_2(\r)$, and so on,
distributing the $N$ atoms among $n_{orb}>1$ orthogonal orbitals.
At the other end to the Gross-Pitaevskii approach lies the situation where {\it each}
boson in the optical lattice resides in a different orbital, i.e., $n_{orb}=N$.
More formally, 
the multi-orbital mean-field wavefunction for $N$ interacting bosons is 
the following {\it single} configuration wavefunction \cite{LA_PLA}
\beq\label{general_MF}
  \Psi(\r_1,\r_2,\ldots,\r_N) = \hat{\cal S} 
 \left\{ \varphi_1(\r_1) \varphi_2(\r_2) \cdots \varphi_N(\r_N) \right\}, 
\eeq 
where $\hat{\cal S}$ is the symmetrization operator.
Note that the Gross-Pitaevskii 
approach is a specific case of Eq.~(\ref{general_MF}) where all orbitals are alike \cite{LA_PLA,LA_ALN_PRA}.
In order to find the ground state of the many-bosonic system 
with the ansatz (\ref{general_MF}) one has to minimize the 
energy $<\Psi|\hat H|\Psi>$ with respect to $\Psi$, Eq.~(\ref{general_MF}).
This results in a set of $n_{orb}$ coupled, non-linear equations for
the $n_{orb}$ orbitals that have to be solved {\it self-consistently}.
To fulfill the variational principle,
 we should search for the energy minimum of $\hat H$ with respect to:
(i) the shape of the {\it self-consistent} orbitals $\varphi_i(\r)$,
(ii) the occupation of each orbital and
(iii) the number $n_{orb}$ of different orbitals.

The multi-orbital mean-field wavefunction (\ref{general_MF})  
has been successfully employed and led us
to the prediction of macroscopic fragmentation of repulsive condensates 
in the ground and excited states, see Ref.~\cite{LA_ALN_PRA}. 
In macroscopic fragmentation, a large number of atoms
reside in a small number of orbitals.
Specifically, two or three orbitals were 
considered in Ref.~\cite{LA_ALN_PRA} within Eq.~(\ref{general_MF}).
What we have found out is that, 
when the number of orbitals is of the order of $N$ (number of atoms),
the multi-orbital ansatz (\ref{general_MF})
is physically very relevant for {\it various} situations of atoms in optical lattices, see below.
At the same time, we can now handle practically any number of atoms and sites in optical lattices.
Specifically here, we employ more than 100 bosons/sites, i.e., 
the number of orbitals is $n_{orb}>100$. 
More details will be reported elsewhere.

In this work, we concentrate on 1D optical lattices,
$V(x)=V_0 \sin^2(kx)$, where $k$ is the wave vector.
As usual, periodic boundary conditions are assumed, thus we employ a super-cell of a large 
number $N_w$ of potential wells (unit cells).
The strength of the inter-particle interaction is expressed via 
the dimensionless parameter $\gamma=m g/\hbar^2 n$,
where $m$ is the mass of the atoms, $n$ is the density, and
$g$ is related to the scattering length and the transverse harmonic 
confinement \cite{Maxim_PRL}.
Optical potential depths, $V_0$, 
and energies will be expressed in terms of the recoil energy, 
$E_R=\hbar^2 k^2/2m$.

We begin our study with 
bosons in a deep optical lattice, $V_0=25 E_R$,
with commensurate filling of one atom per lattice site.
This familiar SF to MI phase transition is to serve as a test tool for 
our multi-orbital mean-field approach. 
The MI phase in this system will be denoted by MI(1).
In practice, we minimized the energy of the Hamiltonian $\hat H$ with respect to
the ansatz (\ref{general_MF}).
For weakly-interacting bosons we found that {\it all} bosons reside in one orbital,
namely, that the Gross-Pitaevskii approach provides the lowest energy.
This state, naturally, corresponds to the SF phase.
For stronger interactions, the situation for the ground state changes completely:
{\it each and every atom} occupies now a different self-consistent, orthogonal orbital!
This, as we shall see below, is the MI(1) phase.
In other words, we employ the ansatz (\ref{general_MF})
with $n_{orb}=N$ equations to describe the MI(1) phase.
In Fig.~1A, we plot the energy per particle $\varepsilon$ of the SF and MI(1) 
states as a function of the dimensionless parameter $\gamma$ for $N_w=102$ sites.
The crossing of the energy curves indicates the critical value
for which the phase transition from the SF to the MF(1) phase occurs.
It is found to be $\gamma_c=0.007777(3)$.
It is instructive 
to translate this value to the ``language'' of the Bose-Hubbard model,
i.e. express 
the corresponding $U/J$ (on-site interaction divided by hopping)
in terms of $V_0/E_R$ and $\gamma$, see \cite{Jaksch1_PRL,Zwerger1_PRL}.
The resulting critical value corresponding to our $\gamma_c$  
is readily found to be $(U/J)_c = 3.855(7) = 2 \cdot 1.927(8)$.
It is very encouraging to find out that, 
although our approach is {\it mean-field} in real space, 
it finds a value which is in excellent agreement with density matrix 
renormalization group and other {\it many-body} calculations of the 1D Bose-Hubbard model,
see Ref.~\cite{Zwerger2_EPL} and references therein.
We remind in this context 
that the {\it mean-field} critical value 
of the 1D Bose-Hubbard model is $(U/J)_c \approx 2 \cdot 5.8$,
rather off this value, 
see \cite{Jaksch1_PRL} and references therein.

How do the corresponding wavefunctions (orbitals) look like?
In Fig.~1B we show the SF phase orbital, i.e., the corresponding Gross-Pitaevskii orbital.
It, of course, extends throughout the optical lattice, representing a coherent state.
In Fig.~1C shown are a few adjacent orbitals describing the MI(1) phase.
We remind that these orbitals are obtained {\it self-consistently}
as the solution of a coupled system of $n_{orb}=102$ non-linear equations, following ansatz (\ref{general_MF}).
Namely, {\it no} preliminary assumptions are made regarding their shape.
Due to the translational symmetry, a set of equivalent orbitals is obtained, 
each centered around one lattice site.
It can be proved that, 
for {\it vanishing} inter-particle interaction, 
these orbitals {\it approach} 
the lowest-band Wannier functions.
This suggests a very appealing physical meaning to the ansatz (\ref{general_MF}).
It describes the MI(1) phase with orbitals that can
be interpreted as {\it boson-dressed Wannier functions}.

With a successful mean-field real-space description of the standard SF to MI transition,
we set in to exploit more advantages of our method.
The next system we considered is the SF to MI(1) phase transition in {\it shallow}
optical lattices (here we took $V_0=E_R$),
i.e., for a system of {\it strongly-interacting} bosons.
For shallow optical lattices the MI(1) orbitals become more diffusive, as can be seen in Fig.~1D,
and extend beyond the next-nearest neighbor sites.
Accordingly, the corresponding spatial density $\rho$ becomes flatter -- compare Figs.~1C and 1D.
Also, notice the {\it negative} values of the orbitals in the nearest neighbor wells
which ensure their orthogonality.
For $V_0=E_R$,
 the value of the SF to MI phase transition has been determined 
to be $\gamma_c=3.4(9)$.
It is instructive to find this mean-field value to agree very well 
with that found by B\"uchler {\it et al.} in the 
limit $V_0 \to 0$ by employing the sine-Gordon problem \cite{Zwerger1_PRL}.

Aiming at exploring more physical situations 
in optical lattices,
we considered another scenario of commensurate filling -- with two atoms per lattice site.
For weak interaction,
the ground state is, of course, the SF phase where all bosons
reside in a single orbital.
As the interaction increases, a phase transition to the 
MI phase, which will be denoted here by MI(2), occurs.
We obtained the MI(2) phase as the ground state of the system by ansatz (\ref{general_MF}).
Technically, this is done by determining the $n_{orb}=N/2$ self-consistent orbitals
minimizing the energy with $2$ bosons per orbital.
As $\gamma$ is further increased,
it passes another critical value,
and we find that the ground state of the system underwent 
a second, new phase transition (see Fig.~2A).
Now, there are still two atoms in each site but they reside in {\it different} orbitals, 
in contrast to the MF(2) phase where they reside in the {\it same} orbital.
We denote this new MI phase by MF(1,1) (for obvious reasons). 
In real space, the two atoms in each site localize at the borders of the site,
leading to minima in the spatial density.

Figs.~2B and 2C present the wavefunctions (orbitals) of the standard and new MI phases,
MI(2) and MI(1,1), respectively.
The appearance of on-site minima in the corresponding spatial density 
is a {\it clear} fingerprint of the new MI phase,
differentiating it from the standard MI(2) quantum phase.
Differences between the standard and the new MI phase should also 
manifest themselves in other observable quantities.
For instance, the momentum profile of the MI(1,1) phase should be flatter with respect to
that of the MI(2) phase. 
Similarly, absorption images resulting from a sudden release of the atoms from the optical lattice 
in the MI(2) phase, where each site houses 2 coherent atoms, 
should have a slightly sharper interference pattern with respect to the MI(1,1), 
where each site houses 2 atoms in orthogonal states.  
The transition of MI(2) to the quantum phase MI(1,1) when the interaction is increased can
be interpreted as an on-site transition to the Tonks-Girardeau regime of the MF(2) phase. 
The results obtained so far are straightforwardly extended to
more MI phases with more atoms per site.
For instance, the MI phase with three atoms per lattice site, 
which we denote by MI(3),
would eventually end up as the new MI phase, MI(1,1,1),
where all three atoms reside in different orbitals, 
and two on-site minima now appear in the spatial density.

So far, we successfully applied the ansatz (\ref{general_MF}) to the {\it ground state} of atoms in optical lattices.
This has been obtained by solving a coupled system of $n_{orb}$ non-linear equations for the orbitals.
It is natural to ask whether the ansatz (\ref{general_MF}) can also provide 
physical information on {\it self-consistent excited} states.
We remind in this respect that ansatz (\ref{general_MF}) boils down
to the Gross-Pitaevskii approach if all orbitals are alike \cite{LA_PLA,LA_ALN_PRA}.
The self-consistent excited states of the latter -- solitons and vortices -- 
are well-known and have been observed for condensates, see, e.g., Ref.~\cite{review_Legget} and references therein.
In optical lattices, the self-consistent excited states are the non-linear Bloch bands,
the lowest of which is quadratic in $k$ and has zero gap \cite{Lattice_Pitaevskii}.
By definition, in the self-consistent excited states of the Gross-Pitaevskii approach 
{\it all} bosons reside in the same, higher-energy orbital.
In contrast, the flexibility of putting atoms in different orbitals suggests
that Eq.~(\ref{general_MF}) can also be employed to describe
many {\it self-consistent excited} states.

In the following, we would like to employ ansatz (\ref{general_MF}) to understand the nature
of the gap in the MI phase of {\it strongly-interacting} atoms in optical lattices.
For this, we recall that for {\it weakly-interacting} bosons
in the MI(1) phase the lowest-in-energy excitation is obtained by 
moving {\it one} boson from its site 
to a neighboring site, see, e.g., Refs.~\cite{IB1_nature,Jaksch1_PRL}.
But, what happens when the interaction between bosons increases, 
entering even deeper to the MI regime?
In that case, the energy of this excited state increases substantially
due to the on-site interaction between the two bosons. 
We, therefore, employed ansatz (\ref{general_MF}) and searched for the low-lying 
self-consistent excited states of the system.
We found many such states, where, e.g.,
two atoms reside in a single delocalized orbital, 
or where two atoms in the {\it same} site occupy different orbitals
localized at the borders of this site (similarly to Fig.~2C). 
For sufficiently strong interaction, the excitation lowest in energy does not accommodate two bosons in the same site or orbital. 
Rather, it emerges as an on-site excitation -- see Fig.~3.
It is instructive to compare the energy of the self-consistent on-site excitation with
the excitation gap of a single well in the harmonic approximation given by $\sqrt{4 V_0 E_R}$ \cite{Jaksch1_PRL}.
For the parameters used here, $\gamma=76.64$, $V_0=25 E_R$ and $N_w=102$,
we find this value to be $0.90 \sqrt{4 V_0 E_R}$,
slightly lower than the bare-well gap.
The difference comes from the strong repulsion between atoms
which lowers the gap in comparison to the interaction-free problem.

In conclusion, weakly- to strongly-interacting cold bosonic atoms 
in deep to shallow optical lattices have been 
described by a multi-orbital mean-field approach in real space,
giving the wavefunction of cold atoms in the lattice.
With it, we described quantum phases and phase transitions of
cold bosonic atoms in 1D optical lattices not accounted for so far.
As the inter-particle interaction is increased, 
on-site excitation becomes the excitation lowest in energy.
The employment of {\it self-consistent mean-field orbitals} has been shown to provide
an accurate value of the SF to MI transition in 1D.
The findings demonstrate the wide potential of
our multi-orbital ansatz for cold bosonic atoms
and motivate concrete studies in higher dimensions.
Finally, the predictions obtained are within reach of (effective) interaction strengths 
presently employed in experiments, where values up to $\gamma=200$ 
have been realized \cite{IB2_nature}.
In addition if needed, Feshbach-resonance techniques 
can be employed to further increase scattering lengths.

\acknowledgments

\noindent
We thank J\"org Schmiedmayer for discussions.


\begin{figure}[ht] 
\includegraphics[width=10cm,angle=-90]{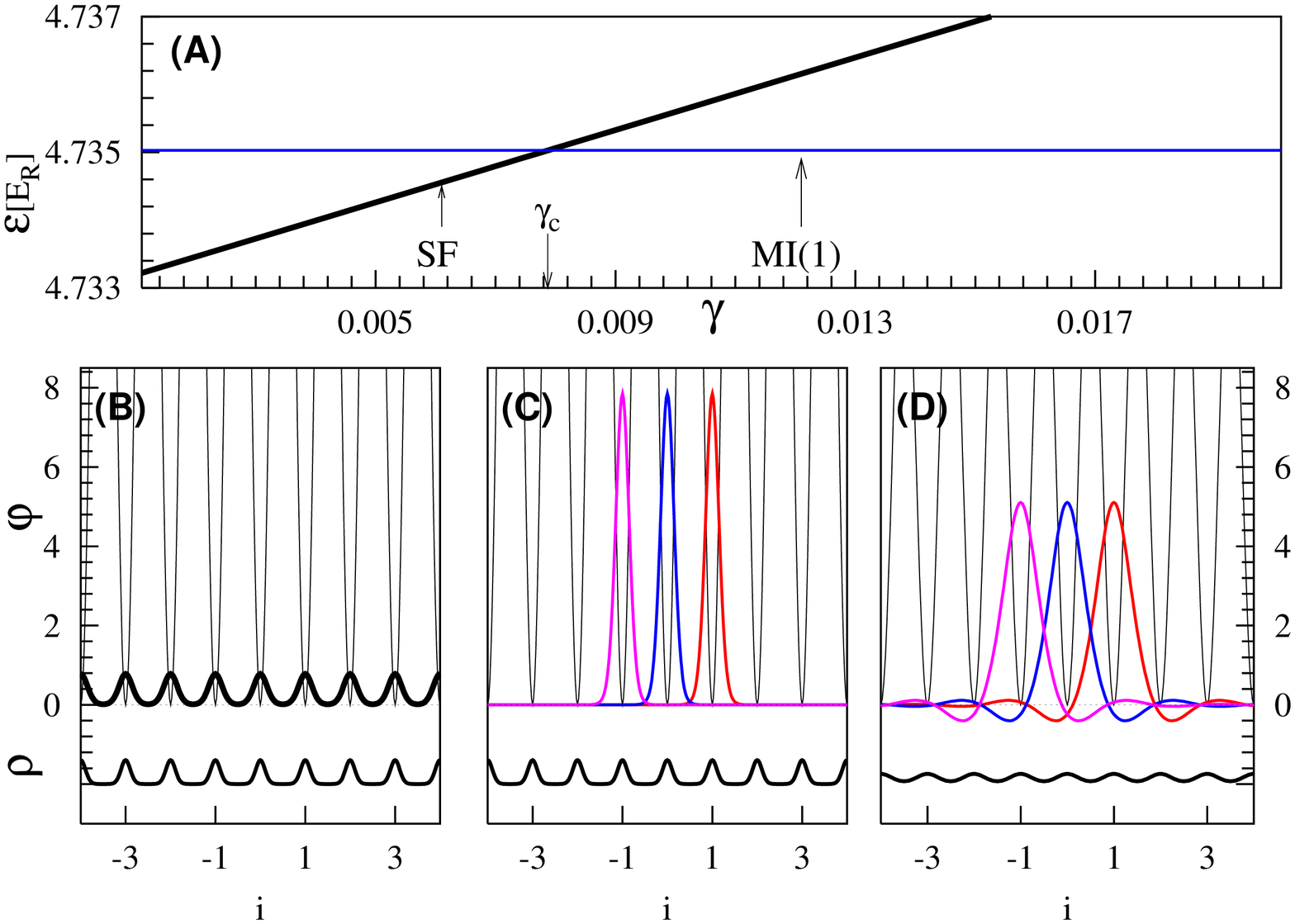}
\caption [kdv]{Quantum phase transition from the SF to MI(1) phase
and corresponding orbitals $\varphi$ and densities $\rho$ (shifted lower curves) for weakly-interacting bosons 
in optical lattice with $V_0=25 E_R$ and $N_w=102$ sites.
Orbitals and densities are normalized (on the segment of length $2\pi$) and are plotted 
against the site index ``i''. The optical lattice is illustrated for guidance by the background sinusoidal curve.
(A) The phase transition is described by the intersection 
of the SF and MI(1) energy per particle curves, $\varepsilon$, which occurs at $\gamma_c=0.007777(3)$.
(B) Orbital and density of the SF phase for $\gamma=0.00776002$.
(C) Orbitals and density of the MI(1) phase for $\gamma=0.00777731$, slightly above $\gamma_c$ (shown are 3 adjacent orbitals).
(D) Orbitals and density of the MI(1) phase for strongly-interacting bosons in shallower optical lattice with $V_0=E_R$ 
for $\gamma=3.491$, slightly above the corresponding $\gamma_c=3.4(9)$ (shown are 3 adjacent orbitals).
}
\end{figure}


\begin{figure}[ht] 
\includegraphics[width=10cm,angle=-90]{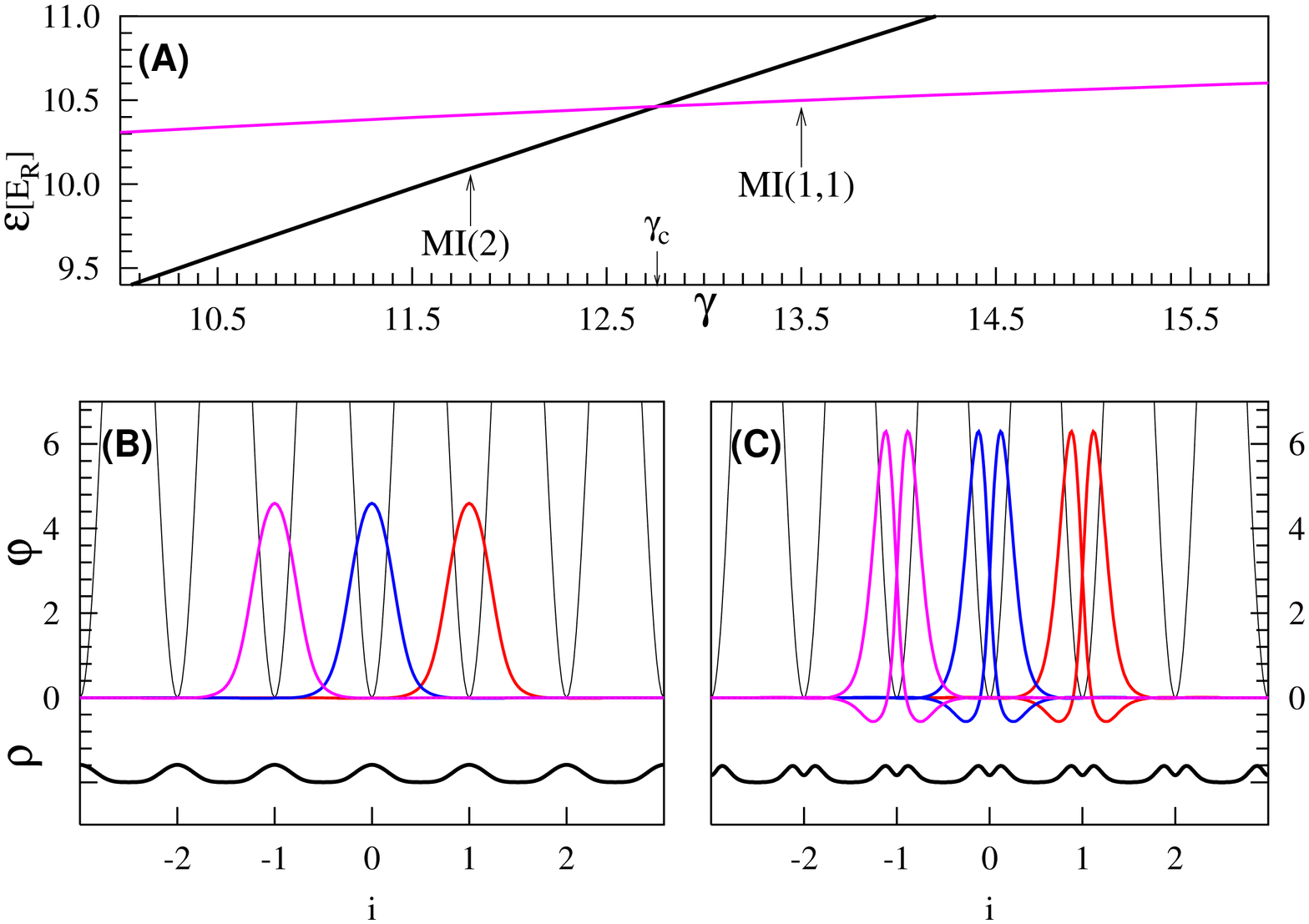}
\caption [kdv] {Quantum phase transition from the MI(2) to MI(1,1) phase
and corresponding orbitals $\varphi$ and densities $\rho$ (shifted lower curves) for weakly-interacting bosons 
in optical lattice with $V_0=25 E_R$ and $N_w=51$ sites.
Orbitals and densities are normalized (on the segment of length $2\pi$) and are plotted 
against the site index ``i''. The optical lattice is illustrated for guidance by the background sinusoidal curve.
(A) The phase transition is described by the intersection of the MI(2) and MI(1,1) 
 energy per particle curves, $\varepsilon$, which occurs at $\gamma_c=12.7(6)$.
(B) Orbitals and density of the MI(2) phase for $\gamma=12.7418$ (shown are 3 adjacent orbitals).
(C) Orbitals and density of the MI(1,1) phase for $\gamma=12.7609$, slightly above $\gamma_c$ (shown are 6 adjacent orbitals).
Notice the on-site minima in the density of the MI(1,1) phase in comparison to the standard MI(2) phase. 
}
\end{figure}

\begin{figure}[ht] 
\includegraphics[width=13cm,angle=-90]{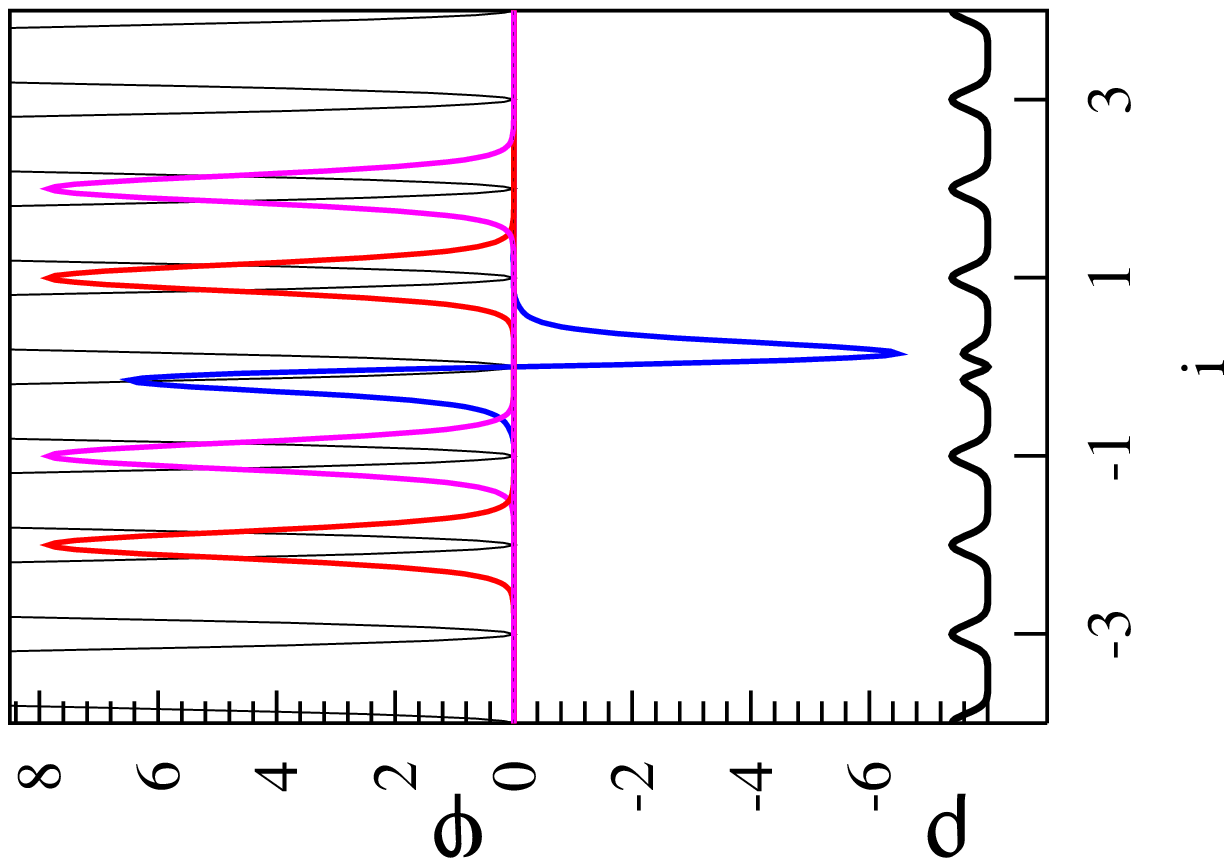}
\caption [kdv]{On-site {\it self-consistent excited} state of the MI(1) phase
in optical lattice with $V_0=25 E_R$ and $N_w=102$ sites for 
{\it strongly-interacting} bosons, $\gamma=76.64$. 
Orbitals and density (shifted lower curve) are normalized (on the segment of length $2\pi$) and are plotted 
against the site index ``i''. The optical lattice is illustrated for guidance by the background sinusoidal curve.
Shown are the excitation site and 4 adjacent orbitals.
For the present $\gamma$, this is the excitation {\it lowest} in energy,
accommodating the energy gap of the MI(1) phase
calculated here to be $0.90 \sqrt{4 V_0 E_R}$.
}
\end{figure}

\ed
\begin{thebibliography}{99}


\bibitem{IB1_nature} M. Greiner, O. Mandel, T. Esslinger, T. W. H\"ansch. and I. Bloch,
Nature (London) {\bf 415}, 39 (2002).


\bibitem{TE1_PRL} T. St\"oferle, H. Moritz, C. Schori, M. K\"ohl, and T. Esslinger,
  Phys. Rev. Lett. {\bf 92}, 130403 (2004).

\bibitem{IB2_nature} B. Paredes, A. Widera, V. Murg, O. Mandel,
S. F\"olling, I. Cirac, G. V. Shlyapnikov, T. W. H\"ansch,
and I. Bloch,
Nature (London) {\bf 429}, 277 (2004).

\bibitem{Fisher_PRB} M. P. A. Fisher, P. B. Weichman, G. Grinstein, and D. S. Fisher,
 Phys. Rev. B {\bf 40}, 546 (1989).

\bibitem{Jaksch1_PRL} D. Jaksch, C. Bruder, J. I. Cirac, C. W. Gardiner, and P. Zoller,
 Phys. Rev. Lett. {\bf 81}, 3108 (1998).

\bibitem{Zwerger1_PRL} H. P. B\"uchler, G. Blatter, and W. Zwerger,
 Phys. Rev. Lett. {\bf 90}, 130401 (2003).

\bibitem{LA_PLA} L. S. Cederbaum and A. I. Streltsov,
 Phys. Lett. A {\bf 318}, 564 (2003).

\bibitem{LA_ALN_PRA} L. S. Cederbaum and A. I. Streltsov,
 Phys. Rev. A {\bf 70}, 023610 (2004); 
 A. I. Streltsov, L. S. Cederbaum, and N. Moiseyev, 
{\it ibid}, 053607 (2004).

\bibitem{Zwerger2_EPL} S. Rapsch, U. Schollw\"ock, and W. Zwerger,
Europhys. Lett. {\bf 46}, 559 (1999).


\bibitem{Maxim_PRL} M. Olshanii,
 Phys. Rev. Lett. {\bf 81}, 938 (1998).

\bibitem{review_Legget} A. J. Leggett, Rev. Mod. Phys. {\bf 73}, 307 (2001).


\bibitem{Lattice_Pitaevskii} M. Kr\"amer, C. Menotti, L. Pitaevskii, and S. Stringari,
Eur. Phys. J. D {\bf 27}, 247 (2003).


\end{thebibliography}
